\documentclass[10pt]{article}
\usepackage{multicol}

\providecommand{\multicols}[1]{}
\usepackage{fullpage}

\newcommand{\average}[1]{\langle #1\rangle}
\newcommand{\eql}[1]{\label{#1}}
\newcommand{\eq}[1]{(\ref{#1})}
\newcommand{\Eq}[1]{Eq.~\eq{#1}}
\newcommand{\VZC}{Ref.~\cite{VanZonCohen03a}}
\newcommand{\ND}{Ref.~\cite{NarayanDhar03}}
\newcommand{\NDs}{Narayan and Dhar's}
\newcommand{\FT}{FT}
\newcommand{\firstFT}{fluctuation theorem~(FT)}

\begin{document}

\title{Comment on ``Reexamination of experimental tests of the
fluctuation theorem'' by Narayan and Dhar}

\author{R. van Zon and E.G.D. Cohen
\vspace{1mm}\\
{\em Rockefeller University, 1230 York Avenue, New York, NY 10021}}

\date{July 11, 2003}

\maketitle

\begin{multicols}{2}

{\em 
Our result in Ref.~\cite{VanZonCohen03a} that, instead of the conventional heat \firstFT, a new \FT\ holds for heat fluctuations for a Brownian particle in a moving confining potential\cite{Wangetal02}, was claimed to be disproved in a very recent preprint by Narayan and Dhar\cite{NarayanDhar03}.  This comment is meant to show that their assertion is not correct. The point is that they formulate their \FT\ differently than we do ours. Effectively, their \FT\ speaks about a physically irrelevant limiting case of our new \FT.  This implies that the two \FT s are not in contradiction with each other.  Furthermore, we point out an incorrect assumption in their derivation.
}

First, let us remark that although we have a \FT\ for all times in \VZC,
we only discuss the infinite time version in this comment.

What we consider as the conventional and the new \FT, and what Narayan
and Dhar in \ND\ consider the \FT\ to be, respectively, can be
formulated as follows.

\newcommand{\sw}{\sigma}
 
\underline{We} \cite{VanZonCohen03a}: $P(Q_\tau=p\sw\tau)$ denotes the
probability that the heat $Q_\tau$ produced in a time interval $\tau$,
has the value $p\sw\tau$, where $\sw$ is the average heat produced per
unit time ($=\average{Q_\tau}/\tau$) in the stationary state over
positive times\cite{GallavottiCohen95a,CohenGallavotti99}.\footnote{In
\VZC, $\sigma$ is called $w$.}

\underline{Narayan and Dhar} \cite{NarayanDhar03}: $P(Q_\tau=A)$
denotes the probability that the heat $Q_\tau$ produced in a time
$\tau$ has the fixed value $A$, where $A$ is independent of $\tau$. We
note that in \ND\ this is formulated in terms of $S$, which is the
entropy production in a time $\tau$, and is given by $S=Q_\tau/T$,
with $T$ the temperature of the fluid in which the Brownian particle
moves.

As a consequence, there are two similarly looking \FT s, but at closer
inspection they are physically radically different.

\underline{We} \cite{VanZonCohen03a} consider the conventional \FT\ in
the form
\begin{equation}
  \lim_{\tau\to\infty} \frac{1}{\beta p\sw\tau} 
  \log\left[\frac{P(Q_\tau=p\sw\tau)}{P(Q_\tau=-p\sw\tau)} \right] 
  = 1
\eql{star}
\end{equation}
for fixed $p$, where $p$ characterizes $Q_\tau$ in units of
$\sw\tau=\average{Q_\tau}$ and $\beta=1/k_BT$.
This is consistent with
Refs.~\cite{GallavottiCohen95a,CohenGallavotti99}.

\underline{Narayan and Dhar} \cite{NarayanDhar03} consider the
conventional \FT\ to be
\begin{equation}
  \lim_{\tau\to\infty} \frac{1}{\beta A} 
  \log\left[\frac{P(Q_\tau=A)}{P(Q_\tau=-A)} \right] 
  = 1
\eql{starstar}
\end{equation}
for fixed, finite $A$.

\NDs\ \FT\ is so formulated --- because of the independence of $A$
from $\tau$ --- that when $\tau$ approaches infinity, so that
$\average{Q_\tau}$ also approaches infinity,
the fluctuation value $A$ remains always finite and therefore much
smaller than $\average{Q_\tau}$. Then only values of $A$ much smaller
than $\average{Q_\tau}$, but no values either near or larger than
$\average{Q_\tau}$, are considered in \NDs\ \FT.  Hence, only an
increasingly restricted range of fluctuation values $A$ are actually
studied in this \FT\ as $\tau$ approaches infinity.

We, in contrast, consider all fluctuation values $A$, because we scale
the value $A$ with $\tau$, so that when $\average{Q_\tau}$ grows
($\sim\tau$), so does $A=p\sw\tau$, as is written explicitly in our
formulation in \Eq{star} for the \FT. A $\tau$-proportional $A$ means
that for a given $\average{Q_\tau}$, {\em any} heat fluctuations
($Q_\tau=p\sw\tau$) can be considered by an appropriate choice of $p$,
even as $\tau$ approaches infinity.\footnote{This issue of scaling is
analogous to using the proper scaling to derive, e.g., the diffusion
equation for the random walk, which requires an $l^2/t$ scaling. Other
scalings (like e.g., $l/t$) would not lead to a physically meaningful
description of diffusion\cite{Kac}.}

So we claim that \Eq{star} is consistent with \Eq{starstar} since a
constant finite $A$ --- as \NDs\ \FT\ uses in \Eq{starstar} --- can
also be obtained from \Eq{star}, by taking the limit $p\to 0$ (when
$\tau\to\infty$). In other words, \Eq{starstar} is a very special case
of \Eq{star} which covers strictly speaking only the value $p=0$,
while \Eq{star} covers all values of $p$.

We will now argue that the FT of Narayan and Dhar in \Eq{starstar} is
also consistent with our new FT found in \VZC.  In \VZC\, the
conventional \FT\ [\Eq{star}] was found to be incorrect as a \FT\ for
heat fluctuations in the system described above, and had to be
replaced by a new \FT. However, for the special region $-1\leq
p\leq1$, this new \FT\ turns out to coincide with the conventional
form of \Eq{star}, which includes $p=0$. Therefore, there is no
contradiction between our new \FT\ and \NDs\ \FT\ in \Eq{starstar}
either.

The above is enough to conclude that our new \FT\ in \VZC\ is not in
contradiction with the result \Eq{starstar} (or their Eq.~(1) in
\ND). Moreover, our theory leading to the new \FT\ in \VZC\ has been
compared in detail with two independent numerical calculations, one, a
numerically carried out inverse Fourier transform explained in \VZC,
and another, a sampling method, which is yet to be
published\cite{VanZonCohen03b}. Both give perfect agreement with each
other as well as with the new \FT. So also from this, we have no reason to
distrust the results in \VZC.

We take this opportunity to address very briefly two other points in
\ND.

Firstly, in their derivation of their (conventional) \FT\ for $Q_\tau$
in their Sec.~III, Narayan and Dhar neglect a term in their Eq.~(17)
which one can neglect for the average $\average{Q_\tau}$, but not for
the fluctuations of $Q_\tau$. In fact, it is precisely this term which
gives rise to exponential rather than Gaussian behavior of the tails
of the distribution $P(Q_\tau)$ --- the existence of which Narayan and
Dhar also notice --- and through these exponential tails, this term
also gives rise to our new \FT, instead of to the conventional one. To
substantiate this, a saddle point
method\cite{VanZonCohen03a,VanZonCohen03b}, involving a complicated
study of the behavior of the Fourier transform of $P(Q_\tau)$,
in particular, its singularities in the complex plane, is needed, which
goes beyond the scope of this comment.

Secondly, Narayan and Dhar seem to suggest that the singularity of the
Langevin equation for vanishing mass $m$ of the particle might be a
problem in the derivation of a \FT. However, setting $m=0$ from the
start in their Eq.~(12) or setting $m=0$ at the end
in their Eqs.~(17) and (19), give the same \FT, i.e., no
singularity is encountered in this derivation with respect to
$m=0$.

\small


\begin{thebibliography}{1}

\bibitem{VanZonCohen03a}
R.~van Zon and E.~G.~D. Cohen,
\newblock An extension of the fluctuation theorem, 2003,
\newblock cond-mat/0305147.

\bibitem{Wangetal02}
G.~M. Wang, E.~M. Sevick, E.~Mittag, D.~J. Searles, and D.~J. Evans,
\newblock Phys. Rev. Lett. {\bf 89}, 050601 (2002).

\bibitem{NarayanDhar03}
O.~Narayan and A.~Dhar,
\newblock Reexamination of experimental tests of the fluctuation theorem, 2003,
\newblock cond-mat/0307148.

\bibitem{GallavottiCohen95a}
G.~Gallavotti and E.~G.~D. Cohen,
\newblock Phys. Rev. Lett. {\bf 74}, 2694 (1995).

\bibitem{CohenGallavotti99}
E.~G.~D. Cohen and G.~Gallavotti,
\newblock J. Stat. Phys. {\bf 96}, 1343 (1999).

\bibitem{VanZonCohen02b}
R.~van Zon and E.~G.~D. Cohen,
\newblock Phys. Rev. {E} {\bf 67}, 046102 (2003),
\newblock (cond-mat/0212311).

\bibitem{Kac}
M.~Kac,
\newblock Am. Math. Monthly {\bf 54} (1947),
\newblock see section 2; This paper was reprinted in 
N.~Wax, editor,
\newblock {\em Selected papers on noise and stochastic processes} (Dover
  Publications Inc., New York, 1954).

\bibitem{VanZonCohen03b}
R.~van Zon and E.~G.~D. Cohen,
\newblock (in preparation).


\end{thebibliography}

\end{multicols}

\end{document}